# Research Report
# Summer Semester 2007

# Static and Dynamic Quality Assurance
# by Aspect Oriented Techniques

by
Christoph Knabe
Prof. of Software Engineering and Programming
http://public.beuth-hochschule.de/~knabe/

at
Technische Fachhochschule Berlin[1]
- University of Applied Sciences -

written 2007-12-08
revised 2010-06-14

---

1   The "Technische Fachhochschule Berlin" was renamed to "Beuth-Hochschule für Technik Berlin" from April 2009.

## Table of Contents

# 1 Introduction

The author was sponsored by a liberation of all university duties for research purposes during the summer semester 2007.

The overall goal of the research project was to create applicable quality assurance patterns for Java software systems using the aspect-oriented programming language extension AspectJ 5 [AspectJ].

The project focussed on static and dynamic measures for quality assurance.

In this report some places, where the author wants to study further, are annotated by „???".

# 2 Research Ways and Results

In this main chapter we will detailedly present the many followed thesises and the few selected decisions, structured according to the domains of the research area. The developed AspectJ 5 aspects are available under [Quality07].

## 2.1 Static quality assurance

### 2.1.1 Mutator convention

In the coding guidelines for LAR AG the author has introduced a classification of attributes and methods with respect to data mutation.

An attribute (Java field) can be declared as mutable by the identifier postfix Mut. All other attributes are considered constant. (As opposed to **final** this means freezing the referenced object, but not the reference).

A method can be declared as mutating by the same identifier postfix Mut. All other methods of the own software system are considered not to mutate any of the attributes of their class.

So the following calls are legal:

```
personMut.setName("Otto");
personMut.promoteMut();
```

but the following calls are illegal:

```
person.setName("Otto");
person.promoteMut();
```

Also the following is illegal, as it replaces an attribute in a non-mutator method.

```
void printSalary(){
    this.name = "Otto";
}
```

#### 2.1.1.1 Checking of the mutator convention

In order to do a static checking of the mutator convention we needed the **declare error** directive of AspectJ. In order to forbid e.g. the call person.promoteMut() we needed the following directive:

```
declare error: call(* *Mut(..)) && target(!(*Mut))
: "Illegal mutator call on an immutable reference";
```

Unfortunately AspectJ does not allow an identifier pattern (as written here) in the **`target`** pointcut, and even during execution time the identifier of the target reference is not available.

In general the AspectJ support for static analysis is very limited. Source code oriented tools as [PMD] showed to be much more suitable for this purpose.

### 2.1.2   Architectural rules

We structured the concerning software system into components, which were each vertically structured into three layers with a strict layering. If we have the three components `user`, `finance`, `production` and a service component `service`, then the resulting package names are shown in the following table. E.g. the user interface layer of the `finance` component should have the package name `finance.ui`.

| **Layer↓** | **Layer abbrev.↓** | `user` | `finance` | `production` | `service` | ←**Component** |
|---|---|---|---|---|---|---|
| User Interface | `ui` | `user.ui` | `finance.ui` | `production.ui` | `service.ui` | |
| Business Logic | `lg` | `user.lg` | `finance.lg` | `production.lg` | `service.lg` | |
| Database Access | `db` | `user.db` | `finance.db` | `production.db` | `service.db` | |

It should be illegal to call the database access layer from the user interface layer directly in order not to avoid the checking of the business rules in the business logic layer. For this purpose we developed a `LayerAspect`, which checks that no method/constructor of the database access layer (`db`) is called from the user interface layer (`ui`) of any product component. The central directive of this aspect looks as follows:

```
declare error
: call(* fb6.*.db.*.*(..)) && within(fb6.*.ui.*)
: "Do not call the db-layer directly";
```

This works quite well, but is not well parameterizable, if we have an a bit more sophisticated layering.

We developed an `ArchitectureAspect`, which prevents the service component from using application components. The central directives are given here with fb6 being the common package name prefix of all components:

```
pointcut inServiceComponent(): within(fb6.service..*);

pointcut callProductComponent(): call(* fb6..*.*(..)) && !call(*
fb6.service..*.*(..));

declare error: callProductComponent() && inServiceComponent()
    : "Do not call a product component from the service component";
```

Unfortunately it was not possible to generalize it in the way, that an application component may call only parts of **itself** and of the service component, but not of other application components. In order to be able to do this „allowing call myself" we needed a more flexible pattern parameterization in the **`declare error`** directive, similar to the capture groups in substitution patterns in the `vi` editor.

## 2.2   Central Exception Reporting

During the research semester the author has decribed a previously developed manual strategy for central exception reporting for the user interface frameworks Struts and Swing. Together with his former student Siamak Haschemi from sd&m the author has developed aspects for doing the same automatically. Together we wrote and published a german article about central exception reporting in the Java Magazin [KnHa07].

## 2.3   Automatically Parameterized Exception Chaining

The framework [MulTEx] gives an easy way to make the best of the known problem with checked Java exceptions. In its usage you should wrap unexpected checked exceptions coming from a lower layer into a `Failure`, which subclasses `RuntimeException` with additional diagnostic information.

A `Failure` contains the causing exception, a message text or key, and actual parameters for the exception message. A typical manual usage of a `Failure` is according to the following scheme:

```
void myMethod(Param1 p1, Param2 p2) throws Precondion1Exc, Precondion2Exc {
    if(!precondition1(p1,p2)){
        throw new Precondion1Exc(p1, p2);
    }
    if(!precondition2(p1,p2)){
        throw new Precondion2Exc(p1, p2);
    }
    try{
        call foreign method 1
        call foreign method 2
        call foreign method 3
        ...
    } catch(Exception ex){
        throw new Failure(
            "Failure executing my method with param1 {0} and param2 {1}"
            , ex, p1, p2
        );
    }
}
```

Here you assure the exception chaining by passing the causing exception into to the `Failure` constructor and you additionally pass other diagnostic parameters.

Often this is good, as you can issue a very appropriate failure message.

But nevertheless, we searched a way to do this wrapping and parameterization automatically. Our goal was to automatically wrap all exceptions, which are not thrown in the *unit* (method/class/package)[2] itself, into a `Failure`, and to enrich its diagnostic capability by all method parameters, and/or all field values as exception parameters.

We took the approach, that exception wrapping and parameterization will be stimulated by a `@WrapDiagnostics` annotation. This can be manually inserted into the source code or it can be created by another aspect, based on layer naming conventions.

A method annotated in this way, should wrap all exceptions, which were not thrown in its unit itself, or were already handled explicitly. Naturally an exception caught explicitly, should not be wrapped inside the **try**-block, as then the **catch**-clause may not apply, as intended by the programmer.

---

2   Further experiments must show, which unit granularity for unwrapped exception propagation is best.

### 2.3.1    Realization by call pointcut

When trying to implement the automatical wrapping using a **call** pointcut, we can intercept all method terminations with an exception by an **after() throwing**(Throwable e): advice.

Sun does not recommend to catch `Error`, but surprisingly an **after() throwing**(Exception e): advice intercepts only method terminations with a checked exception, but not with a `RuntimeException`. May be this is due to the new recursion avoiding restriction of the **declare soft** statement in AspectJ 5, which will no longer soften unchecked exceptions.

In the `WrapUnexpectedExceptionAspect` we tried to determine in the advice body, if there is an explicit handler for this exception surrounding the call joinpoint. An approach to this using the keyword **thisEnclosingJoinPointStaticPart** does not work. Consider the typical code for testing on an expected exception:

```
try {
    StringUtil.dateBefore(null, rightDate);
    fail("NullArgumentException expected");
} catch ( NullArgumentException expected ) {}
```

We do not want, that the `NullArgumentException` will be wrapped into a `Failure`, as then the test will no longer succeed.

Unfortunately we did not find any possibility to determine in the **after throwing** advice for the **call** pointcut, if there is already a specific handler for the intercepted exception.

**thisEnclosingJoinPointStaticPart** does not give access to the **try**-block, but to the execution join point of the surrounding method or initializer block. See the discussion in

http://dev.eclipse.org/mhonarc/lists/aspectj-dev/msg00655.html

There is no pointcut for a **try**-block, too.

### 2.3.2    Realization by execution pointcut

When trying to implement the automatical wrapping using an **execution** pointcut, we can intercept all method terminations with an exception by an
   **after() throwing**(Throwable e):
advice.

<small>As said above: Sun does not recommend to catch `Error`, but surprisingly an **after() throwing**(Exception e): advice intercepts only method terminations with a checked exception, but not with a `RuntimeException`. May be this is due to the new recursion avoiding restriction of the **declare soft** statement in AspectJ 5, which will no longer soften unchecked exceptions.</small>

#### 2.3.2.1    *Wrapping exceptions declared in another unit*

We tried to determine in the advice body, if the terminating exception is declared in another unit than the method we are advising. This can be easily done by inspecting the exception in the join point.

But this approach is not useful when application code is throwing a `multex.Exc` or `multex.Failure`, and not a specialized subclass of it. The possibility to throw an `Exc` or `Failure` directly, has been introduced in order to make exception wrapping and parameterization more simple. The traditional MulTEx approach of declaring a `Failure` subclass for each failable method needs much more coding effort when declaring exceptions. For another solution to this problem see chapter 2.4.

### 2.3.2.2    *Wrapping exceptions, which were thrown in another unit*

A better alternative would be to wrap all exceptions, which were thrown outside of the own unit. If we intercept each exception by an `after() throwing` advice on an execution pointcut, then we could examine its stack trace and by this means determine, where it was thrown.

But, as an exception's stack trace does not contain the location of its `throw` statement, but of its constructor call, the location information can be quite misleading. There are many situations, where exception creating/throwing is delegated to service methods. So we would have to inspect a relevant part of the stack trace. Doing this at each exceptional method completion seems too big an overhead.

Therefor we abandon this approach.

### 2.3.2.3    *Wrapping unspecified exceptions*

So we want to follow another approach. We want to wrap and parameterize all exceptions, which are not specified in the `throws`-clause of a method.

We can by

```
final CodeSignature s = (CodeSignature)thisJoinPointStaticPart.getSignature();
return s.getExceptionTypes();
```

get the classes specified to may be thrown by a method.

We will adopt the convention, that all specified exceptions will be propagated unmodified out of the method. In contrary all unspecified exceptions will be wrapped into a parameterized `Failure` exception.

This is feasible by the `ExecutionDiagnosticsAspect`.

The still unsolved problem is, that this is a runtime aspect, which will not switch off the compiler message `Unhandled exception type` for calling a method, which throws a checked exception. This finally was a motivation for retargeting also the precondition violation base exception class `Exc` to be unchecked from MulTEx 8. See chapter 2.3.2.5. The problem rests nevertheless when calling a traditional Sun method in e.g. package `java.io`. In Eclipse with AJDT this even has a good side: Eclipse optically marks all code locations, where a checked exception must be wrapped into a `Failure` as erroneous, but does not prevent the code from being compiled and executed.

### 2.3.2.4    *Can this be solved by the `declare soft` declaration of AspectJ?*

We would need a way to make the compiler think, that a wrapped exception is no longer in the `throws`-clause, so that its specification will not be propagated through the `throws`-clauses of the complete call hierarchy.

For a similar purpose exists the `declare soft` declaration in AspectJ. By this we can wrap any exception into the unchecked `org.aspectj.lang.SoftException` and wipe it out of the `throws` clause.

This does not really help us, as we want to add diagnostic parameters, too, but the `SoftException` does not have them. As an `after throwing` advice is handled after the softening, we could, though, filter out all softened exceptions, and do one of the following:

a) unwrap and rethrow it, if it was specified in the **throws** clause, or

b) unwrap, and rewrap it into a Failure adding the available diagnostic parameters.

Although these actions would conduct to our goal, we consider this too big an overhead. So we will not follow this way.

May be we want to have this feature enabled independently of wrapping diagnostic parameters.???

### 2.3.2.5    Making the MulTEx Exc-s (precondition violations) unchecked

Now we decided from [MulTEx] 8 upwards to follow the convention, that even precondition violations are unchecked exceptions. Sun itself does not follow this. We do not want to abandon the specification of precondition exceptions in a method header, but we want not to propagate these specifications automatically, although we want to propagate the exceptions (in a wrapped form) automatically.

This will place the complete responsibility for wrapping unspecified exceptions to the runtime aspect. This is done in the `ExecutionDiagnosticsAspect`.

This aspect works well. Only exceptions, that are specified in the **throws**-clause of a method, are propagated unwrappedly. All other exceptions are wrapped, augmented by the argument array of the method, into an unchecked exception `OperationFailure`. This further enables usage of specified exceptions for programming according to the contract model. Additionally this gives very complete diagnostic informations.

When following this way, it is recommendable to do an additional static checking, that will prevent to directly throw an `Exc` by a **throw** statement, if it is not specified in the **throws**-clause. Such a checking would have to be done by a static checker, e.g. by [PMD]. [AspectJ] is not suitable for this purpose. For example the **throw** statement in the following method should not be allowed, as the exception `UsernameNullExc` is not specified in the method header. In contrary specifying an exception makes it propagated unwrappedly.

```
public Person getPersonByUsername(final String username)
throws PersonNotFoundExc, PersonKeyNotUniqueExc {
    if(username==null){
        throw new UsernameNullExc();
    }
    return _getPersonByKey("username", username);
}
```

### 2.3.2.6    Distinguish cause wrapping and parameter capturing

May be it is better to distinguish these 2 goals either by a parameterizable `@WrapDiagnostics` annotation or by providing 2 annotations, which may be both applied to one method independently. ???

### 2.3.2.7    What does `after() throwing(Exception e)` intercept?

Why does'nt it apparently intercept `RuntimeExceptions`???

## 2.4    More Comfortable Exception Declaration

[MulTEx] describes a way, how to declare, specify, throw, handle, and report exceptions. In [MulTEx] each exception is associated with a textual message. The class name of the exception serves as a key for the corresponding message text. The attribute values of the exception object

serve as additional diagnostic information for the message. Direct positional message parameters are stored in an `Object` array.

So usually a well parameterized exception declaration aside with its **throw** statement looks e.g. like this:

```
/**User {0} does not have the right to access file {1}.*/
class FileAccessRightExc extends multex.Exc {
    public FileAccessRightExc(final String username, final File file){
        super(username, file);
    }
}
…
throw new FileAccessRightExc(username, file);
```

In the [MulTEx] version 7 for Java 1.4 we have many constructors in the base class `multex.Exc` with 0 to 9 message parameters. Using the new varargs facility of Java 5 we can shrink this to one constructor with the signature:

```
public Exc(Object... params);
```

Unfortunately, as already criticized in the original MulTEx paper [Knabe00], it is not possible to inherit constructors from a base class in Java. We would like to have our parameterized precondtion violation exceptions declared and thrown very easily, as e.g.:

```
/**User {0} does not have the right to access file {1}.*/
class FileAccessRightExc extends multex.Exc {}
…
throw new FileAccessRightExc(username, file);
```

If we could inherit a varargs constructor of a base class, it would be so simple.

So now we want to study the possibilities to realize this requirement.

## 2.4.1   Realization by AspectJ intertype declaration

AspectJ allows to extend a Java-defined class by specifying additional methods, constructors, attributes, or interfaces. Contrary to normal extending, this does not create a new class, but modifies the extended one.

So we would like to extend each subclass X of `multex.Exc` by a constructor

```
public X(final Object... params){
    super(params);
}
```

Unfortunately an [AspectJ] inter-type declaration does not allow a type pattern to be used to indicate the type(s) to be modified. It allows only one type. So the desirable declaration

```
public aspect InheritExcConstructor {
  /**Extend the direct subclasses of multex.Exc by a parameterized constructor.*/
  public (multex.Exc+ && !multex.Exc).new(Object... params){
    super(params);
  }
}
```

is not compilable.

## 2.4.2   Realization by Aspect Generation

In intertype declarations we can give only a type name, but not a type pattern in order to select the type(s) to be modified. Thus we could solve our problem by generating an intertype declaration for each exception type present in the considered software system.

But we judge this approach as rather brute force. It would probably raise difficult configuration and handling issues. So we will not follow it.

## 2.4.3   Realization by AspectJ `pertypewithin` clause

[AspectJ] allows to declare an aspect with a **pertypewithin** clause giving a type pattern. Then an instance of this aspect is created for each type selected by the type pattern.

So the following code should express our goal:

```
public aspect InheritExcConstructor pertypewithin((multex.Exc+ && !multex.Exc))
{
   /**Extend the direct subclasses of multex.Exc by a parameterized constructor.*/
   public new(Object... params){
     super( params );
   }
}
```

But this is not possible, as well, as the inter-type declaration for the constructor for each subclass of **multex.Exc** is not understood without giving *Type.* before the **new**.

## 2.4.4   Realization by making the base exception class generic

The next way we try is to follow the way of generic collections in Java 5.

We will no longer extend `multex.Exc` by many special-meaning subclasses, but by injecting the special meaning, and thus the error message key by generics into the base class.

This will give us the possibility to

1. distinguish different business exceptions by different parameterizations of `multex.Exc<T>` making them individually catchable.
2. always use the constructor of the base class `multex.Exc` and thus avoid inheriting it.

The usage should look like follows:

```
/**User {0} does not have the right to access file {1}.*/
class FileAccessRight extends multex.MessageKey {}
…
void doAccess(…) throws multex.Exc<FileAccessRight>{
…
    throw new multex.Exc<FileAccessRight>(username, file);
}
…
try{ …
}catch(multex.Exc<FileAccessRight> ex){
    //handle this exception individually
}
```

After detailed investigation this idea showed to be good, but unfortunately not realizable. The Java Language Specification, 3rd ed. states on p. 179 a restriction of the general generics mechanism exactly, where we would need to apply it: „It is a compile-time error if a generic class is a direct or indirect subclass of `Throwable`. ... This restriction is needed since the catch mechanism of the Java virtual machine works only with non-generic classes."

## 2.4.5   Realization by inheriting generic initialization method

The next way we study is, if it is possible to parameterize an exception by an inherited method after creating it by its default constructor.

The code for declaring, specifying and throwing an exception would look like follows:

```java
/**User {0} does not have the right to access file {1}.*/
class FileAccessRightExc extends multex.Exc {}
…
void doAccess(…) throws FileAccessRightExc {
…
    throw new FileAccessRightExc().init(username, file);
}.
```

With Java 1.4 this was not possible, as the inherited method `init` could not be specified to statically return the specific type of object, on wich it is invoked. In Java 1.4 the only possibility to inherit a parameterization method, but to throw the static exception type just created, is as follows:

```java
void doAccess11(…) throws FileAccessRightExc {
…
    final FileAccessRightExc ex = new FileAccessRightExc();
    ex.init(username, file);
    throw ex;
}
```

This notation seemed too heavy. It would not encourage programmers to well-parameterize their exceptions. So this approach was not followed in MulTEx up to version 7.

But with the generic possibilities of Java 5 we should be able to specify the method `Exc.init` with a type parameter `<E>` for defining the result type `E` as follows:

```java
class Exc extends Exception {
    private Object[] params;
    public Exc(){}
    public <E> E init(final Object... params){
        this.params = params;
        return (E)this;
    }
}
```

This allows the following notation for throwing a parameterized exception of a specific static type, here `FileAccessRightExc`:

```java
void doAccess12() throws FileAccessRightExc {
    throw new FileAccessRightExc().<FileAccessRightExc>init(username, file);
}
```

This is correct, but not yet satisfying, as the type of the thrown exception has to be indicated twice. Java cannot infer the type argument from the static type of the primary `new FileAccessRightExc()`, on which the method `init` is invoked.

If we would pass the newly created `FileAccessRightExc` to a generic static `init`-method in class `Exc`, we could infer the type argument, thus avoiding the double indication of the new exception type:

```java
void doAccess2() throws FileAccessRightExc {
    throw Exc.init2(new FileAccessRightExc(), username, file);
```

```
}
```

This notation is working and does not require to doubly indicate the type of the exception to be thrown. Nevertheless it seems to be too heavy. We assume, it will not be broadly accepted by programmers.

### 2.4.6   Realization by invoking generic create method

In the next step we want to unite the creation and the parameterization of the exception into a static generic factory method, which we will call `create`.

The first approach writes as:

```java
void doAccess31() throws FileAccessRightExc {
    throw Exc.<FileAccessRightExc>create(username, file);
}
```

This method invocation is legal Java 5, but we cannot implement this method `create`, as we cannot use a **new** operator on a type parameter.

Thus we have to use the oftenly used form of passing in a class literal, indicating, of which class an object we want to create. The class object will then be used by the compiler to infer the type argument. So we arrive at the form:

```java
void doAccess32() throws FileAccessRightExc {
    throw Exc.create(FileAccessRightExc.class, username, file);
}
```

By using an **import static** declaration for the method `Exc.create` we can still more simplify the creation of a new exception of a definite exception class:

```java
void doAccess32() throws FileAccessRightExc {
    throw create(FileAccessRightExc.class, username, file);
}
```

This notation we judge as follows:
+ naming the thrown exception only once
+ fluently readable
+ throwing an exception is obvious and visible for compiler control flow analysis
- The name `create` is not specific enough.

### 2.4.7   Realization by invoking generic create-and-throw method

In the next step we will study incorporating even the **throw** keyword into the static generic method, which we will call consequently `throwNew`.

The replacement for the **throw new** statement, applying **import static** as before, would look as follows:

```java
void doAccess42() throws FileAccessRightExc {
    throwNew4(FileAccessRightExc.class, username, file);
}
```

This notation seems simple and clear enough, so that programmers can accept to use it. Therefore we present here a complete using example method:

```java
void doAccess43() throws FileAccessRightExc {
    if(!fileAccessAllowed(username, file)){
        throwNew(FileAccessRightExc.class, username, file);
    }
```

```
}
```

This notation we judge as follows:
+ naming the thrown exception only once
+ fluently readable
+ the name `throwNew` is specific enough, to avoid name conflicts
- throwing an exception is not known to the compiler, resulting sometimes in unnecessary warnings

Nevertheless, because usually **throw** and **new** always go together, we prefer this as the style to be used vastly in a software system. The name `throwNew` exactly expresses, what is done inside the method. As this method name for easiness should be used standalone, we need a new utility API class `MultexUtil` in MulTEx, which will hold such methods. It should be imported as follows:

```
import static multex.MultexUtil.*;
```

### 2.4.7.1 Problems of `throwNew` for control flow analysis

In some cases the usage of a method for throwing an exception, instead of a direct **throw** statement causes problems for the control flow analysis of the compiler.

The usual form of checking preconditions as used in the former chapter gives no problem. But e.g. in the following code:

```
if ((allNotifiersWithId == null)||(allNotifiersWithId.size()<1)) {
    throwNew(PersonNotFoundExc.class, keyName, keyValue);
}
final int size = allNotifiersWithId.size();
```

the compiler does not see, that we have checked the reference `allNotifiersWithId` not to be null, and thus gives a warning

> The variable allNotifiersWithId may be null

A still bigger problem are error messages provoked by e.g. the following elegant search algorithm:

```
public static Group getGroupByName(final String name) throws NoSuchGroupExc {
    for (final Group element: KNOWN_GROUPS) {
        if (name.equalsIgnoreCase(element.getName())){
            return element;
        }
    }
    throwNew(NoSuchGroupExc.class, "name", name);
}
```

Here the compiler, not understanding that `throwNew` will certainly throw an exception, complains as follows:
> This method must return a result of type Group

In such cases you have to switch back to the **throw** `create` notation of the former chapter. When adapting the precondition violation exceptions in our `fb6` software system, we found it necessary in 18% of the 45 **throw** statements for `Exc` objects. Although this is not a high percentage, it is already so numerous, that we think about using uniformly only the **throw** `create` notation.

## 2.4.8  Self-checking exception classes

Often people like to define and use global exceptions like the all around

`IllegalArgumentException`. We prefer an exception as a **static** class inner to the throwing class. This holds together data and instructions. If we see, that a specific exception has to be thrown by methods of several classes, then likely not only the exception, but also the checking instructions should be shared!

To illustrate this, we take the ubiqitous argument checking for the String type. We want to throw an `EmptyStringExc`, whenever a **null** or empty String is passed to, where we need real content.

The traditional approach would be, to check for the emptiness in each place, and then condinitonally throw the exception, e.g.:

```java
void doSomethingWithUser(final String username) throws EmptyStringExc {
    if(username==null || username.length()==0){
        throwNew(EmptyStringExc.class, "username");
    }
    …
}
```

Although this exception checking code is not long, it appears quite often and thus is highly redundant. A better way would be to move it to a static method in an utility class `StringUtil`, e.g.:

```java
public static void checkNotEmpty(final String value, final String name)
throws EmptyStringExc
{
    if(value==null || value.length()==0){
        throwNew(EmptyStringExc.class, name);
    }
}
```

But in this way we have an artificial separation of the exception class and the code for checking the exceptional condition and throwing the exception. We now present a solution, which in our opinion is redundancy-free, modular, and universally reusable. We unify the checking code and the exception by providing each exception with a static `check` method, to which we must pass all parameters we need for the checking:

```java
/**String reference {0} is null or has length zero.*/ //message text pattern
class EmptyStringExc extends Exc {

    public static void check(final String value, final String name)
    throws EmptyStringExc
    {
        if(value==null || value.length()==0){
            throwNew(EmptyStringExc.class, name);   //message parameters
        }
    }
}
```

This even does not need an **import static** declaration formerly mentioned, as the `EmptyStringExc` has direct access to the method `throwNew` of its base class `Exc`. The other advantage of this is, that we have the message text pattern (in the first line) and the actual message parameters (commented as such) very near one to another. So we can easily notice a mismatching.

### 2.4.9   Dynamic deriving of the throwing exception class

If we accept the pattern of self-checking exception classes, we still see an eliminable redundancy. The class `EmptyStringExc` mentions in its `throwNew` call its own class name:

```java
throwNew(EmptyStringExc.class, name);
```

By inspecting the actual call stack with help of a `SecurityManager`, we can determine the class of the immediate or the *n*th caller of the actually executed method. This could help us to define a protected method `throwMe` in the base class `Exc`, which will throw a parameterized object of the directly calling class. This will not work for checked exceptions. (But in chapter 2.3.2.5 we already decided to declare even business exceptions as unchecked ones, in order to ensure automatic exception propagation. This will be done from [MulTEx] 8 upwards.)

So we arrive at the following compact style of self-checking exceptions as a general pattern:

```
/**String reference {0} is null or has length zero.*/ //message text pattern
class EmptyStringExc extends Exc {

    public static void check(final String value, final String name)
    throws EmptyStringExc
    {
        if(value==null || value.length()==0){
            throwMe(name);   //message parameters
        }
    }

}
```

Another open question is, if the stack inspection technique used here, will be corrupted by applying AspectJ advices onto a software system. This could be the case, if the call to `throwMe` would be executed inside of an AspectJ-generated method. Fortunately up to now we do not have any indications for this. ???

## 2.4.10   Realization by using enum constants as message keys

The next way we try is to use Java 5 **enum** constants as message keys, as they are, like class names, automatically unique in a Java software system. So we could spare many little exception classes, which have the only purpose to serve as a message key.

This presumably works well and uses less resources. Each `Class` object in Java 5 has 18 reference and 2 **int** fields, which occupy therefore on a 32-Bit-platform about at least 80 bytes, even if the class does not have any members. In contrary an **enum** type is only one class, and each **enum** constant occupies less space in memory than a complete class.

But the advantage of using static member classes over **enum** constants as message keys lies in the fact, that we can place the message key along with its reference message text very near to the **throw** statement. This way the parameter count of the **throw** statement and the message text can be easily held consistent.

## *2.5   More comfortable Failure declaration*

Now we try to make the declaration and throwing of a `Failure` more comfortable. There are many similarities between a `Failure` and an `Exc` declaration, but also some significant differences.

Usually a well parameterized `Failure` declaration aside with its **throw** statement looks e.g. like this:

```
/**Failure loading object of class {0} with id {1}.*/
class LoadObjectFailure extends multex.Failure {
    public LoadObjectFailure(
        final Throwable cause, final Class aClass, final Long id
```

```
    ){
        super(cause, aClass.getName(), id);
    }
}
…
try{
    ...
}catch(Exception ex){
    throw new LoadObjectFailure(ex, aClass, id);
}
```

In order to avoid the heavy declaration of the `LoadObjectFailure` we can follow the simplifications used for the `Exc` classes. The result would be (moving the `.getName()` to the thrower):

```
/**Failure loading object of class {0} with id {1}.*/
class LoadObjectFailure extends multex.Failure {}
…
try{
    ...
}catch(Exception ex){
    throwNew(LoadObjectFailure.class, ex, aClass.getName(), id);
}
```

As mentioned earlier on page 13, the usage of a method to throw an exception causes problems for the control and data flow analysis. Because `Failures` are typically thrown at the textual end of a method body, the compiler does not understand, that the method returns a result in all cases of normal completion.

So we had to switch back to the **throw** create notation in 47% of the 45 statements throwing a `Failure` object. This is a much higher percentage than it was with throwing `Exc` objects.

So we decided, for simplicity of learning, to support in future only the **throw** create notation.

## *2.6   Not-null-Checking*

The author developed a `NullableAspect` and a `@Nullable` annotation. The aspect checks firstly, that only parameters annotated as `@Nullable` will accept a **null** value as argument. Secondly it checks, that only methods annotated as `@Nullable` can return a **null** result.

Violations of these precondition or postcondition are answered by throwing a specific `Failure`, with an explanative message text. For example the method

```
public <T extends DbObject> List<T> getObjects(
    String             oqlQuery,
    @Nullable Object... args
)
```

will accept **null** arguments only for the parameter `args`, as it is annotated as such. A **null** argument for the parameter `oqlQuery` will be answered by a `Failure` exception with the message text:

> Argument "oqlQuery" of executable "Castor.getObjects(String, Object...)" is null, although not annotated as @Nullable

The checking done by this aspect needs quite a lot of runtime resources, as it is applied to each method/constructor execution. But it is not dangerous to switch off this checking, as the same code will then run into a `NullPointerException`. The main purpose of this checking is to provide better diagnostics.

# 3  Conclusions

## 3.1  Static Quality Assurance

AspectJ showed to be not very suitable for static quality assurance.

You can define own compiler warnings and errors, but the patterns, where to apply them, are too restricted.

So you cannot state general conventions, e.g. for layering or for architectural dependencies, which will be checked by AspectJ. But you would have to state each individual illegal dependency as a **declare** directive.

So we conclude, that you can use [AspectJ] for some quality tests, if you anyhow are already using it. But it is not worth to be introduced for static code checking. Other more specialized tools, which make use of full access to the source code, and not only to the byte code, are much more suitable.

## 3.2  Dynamic Quality Assurance

The developed `NullableAspect` is a very good help in rapid diagnosing of illegal **null** values during execution of a program or a test suite.

The developed generic `create` method for easy declaration and creation of parameterized exceptions is usable very well and will probably boost up the usage of well-parameterized exceptions in application systems.

The developed `ExecutionDiagnosticsAspect` for auto-wrapping of unspecified exceptions along with actual method parameters is quite usable, but should be evaluated further. Especially we must gain experience, in which kind of methods we should activate it by the `@WrapDiagnostics` annotation.